\documentstyle[own_sngl]{article}
\def\simless{\mathbin{\lower 3pt\hbox
   {$\rlap{\raise 5pt\hbox{$\char'074$}}\mathchar"7218$}}}   
\def\simgreat{\mathbin{\lower 3pt\hbox
   {$\rlap{\raise 5pt\hbox{$\char'076$}}\mathchar"7218$}}}   
\slugcomment{{\em  MNRAS, in press, 10.03.98}}
\lefthead{Kroupa, P.}
\righthead{Kinematics and Binary Statistics}

\begin{document}

\title{On the Binary Properties and the Spatial and Kinematical 
Distribution of Young Stars}
\author{Pavel Kroupa\\
{\small Institut f\"{u}r  Theoretische Astrophysik,
Universit\"{a}t Heidelberg,\\
Tiergartenstr. 15, D-69121~Heidelberg,
Germany}\\
{e-mail: pavel@ita.uni-heidelberg.de}
}

\begin{abstract}
The effects which star cluster concentration and binarity have on
observable parameters, that characterise the dynamical state of a
population of stars after their birth aggregate dissolves, are
investigated. To this end, the correlations between ejection velocity,
binary proportion, mean system mass, binary orbital period and mass
ratio are quantified for simulated aggregates. These consist of a few
hundred~low-mass binary and single stars, and have half-mass radii in
the range~2.5 to 0.08~pc.  The primordial binary-star population has a
period distribution similar to that observed in Taurus-Auriga for
pre-main sequence binaries.  The findings presented here are useful
for interpreting correlations between relative locations and proper
motions, binary properties and masses of young stellar systems within
and surrounding star forming regions, and of stellar systems escaping
from Galactic clusters.

For the low-concentration binary-rich aggregates, the proportion of
binaries decreases monotonically as a function of increasing ejection
velocity after aggregate dissolution, as expected. However, this is
not the case for initially highly concentrated binary-rich
aggregates. The reason for this difference is the interplay between
the disruption of binary systems and the initial depth of the
potential well from which the stellar systems escape.

After aggregate dissolution, a slowly expanding remnant population
remains. It can have a high binary proportion (80~per cent) with a
high mean system mass, or a low binary proportion (less than about
20~per cent) with a low mean system mass, if it was born in a low- or
a high-concentration aggregate, respectively.  It follows that
adjacent regions on the sky near some star-forming clouds can have
young populations with different binary proportions and different mass
functions, even if the binary proportion at birth and the initial mass
function (IMF) were the same.

Binary systems that are ejected from the aggregate tend to be massive,
and their mass ratio tends to be biased towards higher values.  The
mean system mass is approximately independent of ejection velocity
between~2 and~30~km/s. Dynamical ejection from binary-rich aggregates
adds, within~10~Myr, relatively massive systems to regions as far
as~300~pc from active star-forming centres.  Long-period systems
cannot survive accelerations to high velocities. The present
experiments show that a long-period ($>10^4$~d) binary system with a
large velocity ($>30$~km/s) cannot be ejected from an aggregate.  If
such young systems exist, then they will have been born in
high-velocity clouds.
\end{abstract}


\section{Introduction}
\label{sec:intro}
\noindent Stellar systems (i.e. single or multiple stars) form in
groups. The dynamical processes within these alter the properties of
the young systems when they leave the site where they formed.  The
dynamical properties of a stellar system are its mass (i.e. luminosity
if age is known), the multiplicity and the orbital parameters if it is
a multiple system.  The distribution of velocities of young stars
emanating from star-forming centres (i.e. the kinematical signature of
star formation) will also be affected by the dynamical interactions
within the young groups. Both, the distribution of {\it dynamical
properties} and the {\it kinematical signature of star formation}
bear an imprint of the dynamical configuration at the time when the
stellar group was born.

Star formation in Taurus-Auriga gave birth to aggregates with sizes of
roughly 0.5--1~pc consisting of about 20--50 stars.  It is now well
established that most stars form in binary systems in Taurus-Auriga
(e.g. K\"ohler \& Leinert 1998).  The same appears to hold true in
other star-forming regions (Ghez et al. 1997).  Embedded clusters may
also have a binary proportion that is higher than in the Galactic
field (Padgett, Strom \& Ghez 1997).  In the Trapezium cluster, which
is a very dense embedded cluster and probably less than 1~Myr old,
Prosser et al. (1994) find a binary proportion that is at least as
large as in the Galactic field. In the cluster core, Petr et
al. (1998) observe, for low-mass stars, a binary proportion similar to
the Galactic field, and smaller by about a factor of three than the
binary proportion in Taurus-Auriga.  These findings are particularly
interesting, because binary destruction is expected to be efficient in
such an environment.  A review of pre-main sequence binary stars, and
their relation to Galactic field systems, is provided by Mathieu
(1994, see also Kroupa 1995a, Simon et al. 1995).  Young Galactic
clusters also contain binary systems. The particularly well studied
Pleiades and Praesepe clusters have binary proportions of 40--50~per
cent, for systems of spectral type earlier than K0 (Raboud \&
Mermilliod 1998a, 1998b).

There exists thus evidence that the formation of binary systems may be
by far the dominant star-formation mode in both loose groups and
highly concentrated embedded clusters, some of which may evolve to
bound Galactic clusters. The term {\it aggregates} is used henceforth
to mean loose groups or embedded clusters of more than 10~stars.

If stars form predominantly in aggregates of binary systems, then the
kinetic energy distribution after aggregate dissolution should be
enhanced at high energies, when compared to dissolved aggregates of
single stars, because binary star binding energy can transform into
kinetic energy (Heggie 1975, Hills 1975, Hut 1983).  Large
accelerations are destructive to binary systems, so that the
proportion of binaries should be a decreasing function of increasing
final kinetic energy.  Additionally, different initial aggregate
concentrations lead to different final binary proportions and
kinematical signatures, as will be shown here.  This is also true
under the extreme assumption that {\it all} stars always form in
binary systems with the same initial dynamical properties. This
dynamical mechanism of producing variations in binary proportion and
associated dynamical properties stands in contrast to a possible
variation of these parameters determined by the star-formation
process. Durisen \& Sterzik (1994) make the interesting point that the
binary proportion may be smaller in molecular clouds with a higher
temperature than in lower-temperature clouds.

It is important to study the signatures that arise from purely
dynamical interactions in stellar groups, for a comparison with
outcomes from usually less well-understood alternative scenarios.  In
a study of the large-scale distribution of young stars around active
star-forming regions, Sterzik \& Durisen (1995) find that the
dynamical decay of small stellar groups can lead to sufficiently large
velocities to populate large areas on the sky with young stars, so
that these need not have formed near their observed location. They
find that special initial dynamical configurations of the stars
(e.g. cold thin strings) lead to enhanced production of ejected
stars. Initial decay of cold sub-groups within larger complexes also
has this effect (Aarseth \& Hills 1972), and scattering of proto-stars
on cloud clumps during an even earlier dynamical phase may likewise
eject very young low-mass stars (Gorti \& Bhatt 1996).  However, the
number of ejected stars cannot account for the observed number of
widely distributed young stars (Feigelson 1996).  The evolution of
circum-stellar discs around stars ejected from small stellar groups is
studied by Armitage \& Clarke (1997), and McDonald \& Clarke (1995)
show that the presence of circum-stellar material in small
proto-stellar groups increases the number of binaries formed and
randomises the mass-ratio distribution.  That the number of
dynamically ejected stars is increased significantly in binary-rich
stellar aggregates, when compared to clusters consisting initially
only of single stars, is shown by Kroupa (1995c).  These simulations
show that a mass-ratio distribution produced by randomly associating
masses from the IMF, decays to the observed distribution for G-dwarf
binaries, if most stars form in aggregates similar to observed
embedded clusters.  Also, initially more concentrated aggregates
produce more stars with a high ejection velocity, the maximum of which
increases with decreasing cluster radius.  De la Fuente Marcos (1997)
investigates the dependence on cluster richness, and finds that the
mean ejection velocity increases for more initially populous clusters.
Ejection velocities larger than a few hundred~km/s can be achieved in
young star clusters containing massive primordial binaries (Leonard \&
Duncan 1990). This may explain the location of OB stars far from
active star-forming sites. Leonard (1991) finds, on the basis of many
scattering experiments, that the maximum ejection velocity is of the
order of the escape velocity from the stellar surface of the most
massive star. If its mass is $60\,M_\odot$, then a similar star can
attain an ejection velocity of up to 700~km/s. A low-mass star may
find itself fleeing with a velocity of up to 1400~km/s, after a
surface-grazing encounter with such a star. A critical discussion of
the possible origin of runaway OB stars is provided by Leonard
(1995). He stresses that collisions of two stars during binary-binary
interactions can produce runaway OB stars with very similar properties
as in the alternative scenario, in which such stars result from a
supernova explosion in close binary systems. An interesting and
insightful discussion of the implications of the binary properties of
runaway OB stars for the dynamical configuration of massive stars at
birth is to be found in Clarke \& Pringle (1992).

In this paper, the correlations between stellar velocity, system mass
and binary proportion that arise from aggregates with different
initial concentration and consisting initially either of 400~single
stars or of 200~binary systems, is studied.  The resulting correlations
are useful for interpreting the properties and distribution of young
stars near and in star forming regions (see for example Brandner et
al. 1996, Feigelson 1996, Frink et al. 1997).

In Section~\ref{sec:method} the assumptions, simulations and
definitions are described.  The results are presented in
Section~\ref{sec:results}, and Section~\ref{sec:conclusions} contains
the conclusions.

\section{Method}
\label{sec:method}
\noindent
The initial conditions and numerical method are described in
Section~\ref{subsec:assumptions}, and the data analysis is outlined in
Section~\ref{subsec:observables}.

\subsection{Assumptions}
\label{subsec:assumptions}
\noindent
$N_{\rm bin}=200$ binary systems are distributed in virial equilibrium
according to the Plummer density law, with initial half mass radii
$R_{0.5}=2.53, 0.77, 0.25, 0.077$~pc. These approximately span the
region of parameter space similar to distributed (e.g. Taurus-Auriga)
and very tightly clustered (e.g. Trapezium cluster) star formation.
The clusters have zero initial centre-of-mass velocities in the local
standard of rest.  The $R_{0.5}=0.8$~pc aggregate is especially
interesting, because inverse dynamical population synthesis (Kroupa
1995a,b) suggests that it may be representative of the dynamical
structures in which most stars form (compare with Lada \& Lada 1991).

While the mechanism of binary system formation cannot be specified in
detail, the assumption that the majority of all stars form in binaries
is supported by observational evidence (see review by Mathieu 1994),
and by recent advances in the theory of star formation (for reviews
see Boss 1995, Clarke 1996). However, theory cannot, at present,
constrain the early dynamical properties of stellar systems.  The
interesting suggestion has been made (Durisen \& Sterzik 1994) that
cloud temperature may influence the binary proportion, such that it
may be lower in dense embedded clusters.  For comparison with the
binary rich aggregates, $N_{\rm sing}=400$ single stars are
distributed in aggregates, initially with $R_{0.5}=0.25, 0.077$~pc.

The initial velocity dispersion, $\sigma$, and escape velocity from
the centre of the aggregates, $v_{\rm esc}=\sqrt{2\left|\phi\right|}$
($\phi$ is the Plummer potential at the origin), are:
$\sigma=0.3$~km.s, $v_{\rm esc}=0.77$~km/s ($R_{0.5}=2.5$~pc),
$\sigma=0.5$~km/s, $v_{\rm esc}=1.4$~km/s ($R_{0.5}=0.8$~pc),
$\sigma=0.9$~km/s, $v_{\rm esc}=2.4$~km/s ($R_{0.5}=0.25$~pc), and
$\sigma=1.7$~km/s, $v_{\rm esc}=4.4$~km/s ($R_{0.5}=0.08$~pc).  Other
physical parameters are listed in table~1 of Kroupa (1995a).
Aarseth's NBODY5 programme (Aarseth 1994) is employed for the N-body
simulation of the dynamical evolution of each aggregate in a standard
Galactic tidal field.

In order to simplify the computational burden, the stars are treated
as point particles and stellar evolution is neglected.  The assumption
of virial equilibrium is the simplest case, and implies that the
results presented here are strictly only applicable to escaping stars
from Galactic clusters. The present results can, however, also be used
as guidelines of the type of correlations one might find after
embedded clusters dissolve. An explicit formulation of this problem
requires treatment of gas expulsion, and thus the introduction of
additional ill-defined parameters.  As gas expulsion is not treated
here, the results are representative of star formation with high
efficiency, i.e. aggregates with low residual gas content.  In the
alternative case of a low star formation efficiency, the major effect
gas expulsion has, is a shortening of the time-scale during which the
dynamical evolution occurs. This can be compensated for by a reduction
of $R_{0.5}$, in order to obtain the same effective dynamics
(section~6.4 in Kroupa 1995a). The initial $v_{\rm esc}$ is then
larger.

Stellar masses, $m$, with $0.1\,M_\odot\le m\le 1.1\,M_\odot$, are
obtained from the IMF: $\xi(m)\propto m^{-\alpha_i}$, $\alpha_1=1.3$
for $0.08\,M_\odot \le m<0.5\,M_\odot$, $\alpha_2=2.2$ for
$0.5\,M_\odot \le m<1.0\,M_\odot$ (Kroupa, Tout \& Gilmore 1993), and
$\alpha_3=2.7$ for $1.0\,M_\odot\le m$ (Scalo 1986), where
$\xi(m)\,dm$ is the number of stars with masses in the range $m$ to
$m+dm$.  The mean stellar mass is $0.32\,M_\odot$, and the mass of
each aggregate amounts to $M_{\rm tot}=128\,M_\odot$. Adopting for the
mass of B~stars $6-18\,M_\odot$, each should have associated with it
280~stars with mass in the range $0.08-1\,M_\odot$.  The maximum
ejection velocity that can be achieved is thus limited to about
600~km/s for $0.1~M_\odot$ stars, and about 300~km/s for G-dwarfs
(Leonard 1991).

The main-sequence mass-ratio distribution for G-dwarf binaries
(Duquennoy \& Mayor 1991) is not consistent with random pairing from
the IMF, but may be derived from this assumption if most stars form in
embedded clusters (Kroupa 1995a,b).  In accordance with this result,
and the evidence presented by Leinert et al. (1993), stellar masses
are combined at random to generate the initial binary-star population.
Special care must be taken when interpreting an observed mass-ratio
distribution, as it can be affected significantly by even simple
observational bias (Trimble 1990, Tout 1991).

Binary systems must arrive on the birth-line with eccentricities
approximately dynamically relaxed, because subsequent thermalisation
in the stellar aggregate is not efficient enough to produce such a
distribution (Kroupa 1995b). This is because the cross-section for a
significant change in eccentricity decreases very steeply with increasing
distance of closest approach of a perturber (Heggie \& Rasio 1996).
Consequently, the initial eccentricity distribution is taken to be
dynamically relaxed. The results are not sensitive to this assumption,
however.

An initial period distribution that is consistent with the
observational data for young binaries is used.  The orbital periods,
$P$ (in days), form a flat distribution, $f_{\rm P}({\rm
log}_{10}P)=\left[{\rm log}_{10}(P_{\rm max}) - {\rm log}_{10}(P_{\rm
min})\right]^{-1}$ (equation~3 in Kroupa 1995a), with log$_{10}P_{\rm
min}=3$, log$_{10}P_{\rm max}=7.5$ and $P_{\rm min}\le P\le P_{\rm
max}$.  

For each binary and single-star aggregate, $N_{\rm run}=5$ and~3
simulations, respectively, are carried through.

\subsection{The observables}
\label{subsec:observables}
\noindent
All results quoted here are averages of $N_{\rm run}$ simulations that
are evaluated after 1~Gyr, i.e. after the aggregates have dissolved.
Aggregate dissolution occurs after $700\pm130$~Myr in all cases, when
the number of stars in a volume with a radius of 2~pc, that is centred
on the density maximum of the cluster, has decayed to~3 of less.

The velocity-dependent binary proportion is

\begin{equation}
f_v = {N_{{\rm bin},v}\over(N_{{\rm sing},v}+N_{{\rm bin},v})},
\end{equation}

\noindent 
where $N_{{\rm sing},v}$ and $N_{{\rm bin},v}$ are the number of
single-star and binary systems, respectively, in a velocity interval
$v$ to $v+\Delta v$ relative to the local standard of rest. The
binary proportion in some sub-domain, which may, for example, be the
period range or spatial region accessible to the observer, is
$f=N_{\rm bin}/(N_{\rm sing}+N_{\rm bin})$, where $N_{\rm sing}$ and
$N_{\rm bin}$ are the number of single and binary systems,
respectively, in the sub-domain.  Similarly, $f_{\rm tot}$ is the
binary proportion of the entire population.

The mean system mass in the velocity interval is

\begin{equation}
<m>_v={M_v\over(N_{{\rm sing},v}+N_{{\rm bin},v})},  
\end{equation}

\noindent 
where $M_v$ is the total stellar mass in the velocity
interval. Initially, i.e. at $t=0$, $f_v=1$ and $<m>_v=0.64\,M_\odot$
independent of $v$ for the binary-star aggregates, and $f_v=0$ with
$<m>_v=0.32\,M_\odot$ for the single-star aggregates.

The relative proportion of systems in a velocity interval is

\begin{equation}
h_v = {(N_{{\rm sing},v} + N_{{\rm bin},v}) \over (N_{\rm sing,tot} + N_{\rm
bin,tot})}.
\end{equation}

\noindent
Note that this is $f_v$ in Kroupa (1995c). 

The circular orbital velocity, $v_{\rm orb}$ [km/s], of a binary star
with system mass, $m_{\rm sys}$ [$M_\odot$], and orbital period, $P$
[days], is

\begin{equation}
{\rm log}_{10}P = 6.986 + {\rm log}_{10}m_{\rm sys} - 3\,{\rm
log}_{10}v_{\rm orb}.
\end{equation} 

\noindent
The primordial binary population used here has a maximum $v_{\rm
orb}=27.7$~km/s (for $P=10^3$~d, $m_{\rm sys}=2.2\,M_\odot$) and a
minimum $v_{\rm orb}=0.39$~km/s ($P=10^{7.5}$~d, $m_{\rm
sys}=0.2\,M_\odot$). 

Finally, a stellar system that is {\it ejected} has a final
(i.e. evaluated after 1~Gyr) velocity $v\ge2$~km/s. 

\section{Results}
\label{sec:results}
\noindent
Long distance encounters between systems (two-body relaxation) and in
addition scattering of systems on the non-uniform background potential
(collective effects) dominates the dynamical evolution of the
aggregates. Relatively energetic, stochastically occurring encounters
between stellar systems can lead to the ionization of binary stars and
to the acceleration of a system to escape velocity from the aggregate.

If the mean kinetic energy of the population is ${\overline E_{\rm
kin}}$ and the binding energy of a binary is $-E_{\rm
bin}=-G\,m_1\,m_2/(2\,a)$, where $m_i$ and $a$ are the masses of the
components and the semi-major axis, respectively, then a binary is
termed {\it hard} if $E_{\rm bin}/{\overline E_{\rm kin}}>1$.  Hard
binaries are likely to gain binding energy, i.e. to {\it harden}
(Heggie 1975, Hills 1975), in which case the perturber can be
accelerated to escape velocity. The resulting hardened binary suffers
a recoil which may be sufficient to also expel it from the aggregate.

These processes change the distributions with velocity of the number
of systems, $h_v$, of the binary star proportion, $f_v$, of the mean
system mass, $<m>_v$.  Also, the correlations between binary-star
binding energy, $E_{\rm bin}$, and kinetic energy, $E_{\rm kin}$, and
between the velocity, $v$, and orbital period, system mass and mass
ratio, evolve. The distributions that emerge after aggregate
dissolution thus contain information about the initial dynamical
configuration, but care must be taken in interpreting distribution
data. This is the subject of the present section.

\subsection{Distribution of velocities}
\label{subsec:veldistr}
\noindent
In Fig.~\ref{fig:binprop1}, $h_v$, $f_v$ and $<m>_v$ are plotted as a
function of velocity for the binary star aggregates with $R_{0.5}=2.5,
0.8, 0.25, 0.08$ pc. Fig.~\ref{fig:binprop2} contains the same
information for the two single star aggregates.  In
Table~\ref{table:vdist}, column~1 contains the centre of each
logarithmic velocity bin. Columns~2--7 list, for each aggregate, the
fraction, $h_v$, of systems per logarithmic velocity bin.

After aggregate dissolution, most systems have a velocity near
0.35~km/s (Figs.~\ref{fig:binprop1} and~\ref{fig:binprop2}).  A slight
shift in the maximum of $h_v$ towards smaller $v$ with decreasing
$R_{0.5}$ comes about, because systems have to overcome the initial
aggregate potential before escaping.  The cooling is much more
pronounced in the absence of binary star heating
(Fig.~\ref{fig:binprop2}), and during the first few cluster crossing
times. Later, the aggregate expands to fill its tidal radius and
looses memory of its initial concentration. Initially very
concentrated aggregates, and those with large $R_{0.5}$, have an
indistinguishable life-time (Kroupa 1995c). The binary star
population, however, retains this memory (Kroupa 1995a).

For aggregates with initially smaller $R_{0.5}$, an increase in the
proportion of systems with $v>2$~km/s results.  Comparison of
Figs.~\ref{fig:binprop1} and~\ref{fig:binprop2} shows that, for the
same $R_{0.5}$, aggregates initially with a high proportion of binary
systems have significantly larger $h_v$ at $v>2$~km/s, than
aggregates that consist initially only of single stars.  A high
proportion of primordial binary systems thus increases the percentage
of ejected systems.

\subsection{Binary proportion}
\label{subsec:binprop}
\noindent
High velocity systems are expected to be primarily single stars,
because only relatively hard binaries can survive the large
accelerations during the encounter, as is also stressed by Sterzik \&
Durisen (1995).  This is borne out for the $R_{0.5}=2.5,0.8,0.25$~pc
aggregates (Fig.~\ref{fig:binprop1}), and also in the simulations
reported by Leonard \& Duncan (1990, their figs.~4 and~5) and also
fig.~9 in Kroupa (1995c).

Dynamical evolution is quiescent in star forming regions where the
stellar systems freeze out of the gas in low-density aggregates. The
$R_{0.5}=2.5$~pc aggregate approximates this situation.  In this
aggregate, a binary with $m_{\rm sys}=0.2\,M_\odot$ and $P=10^{7.5}$~d
has $v_{\rm orb}=0.39$~km/s, which is comparable to the velocity
dispersion. The entire binary population is therefore hard, and indeed
most binaries survive cluster evolution.  From such aggregates results
a high ($f_v>0.8$) proportion of binaries for systems with
approximately $v<1$~km/s.  Only 2.6~per cent of the systems end up
with $v>5$~km/s (Table~\ref{table:vdist}), and these have a binary
proportion of approximately $f_v<0.1$ (Fig.~\ref{fig:binprop1}).

For the $R_{0.5}=0.8$~pc aggregate, a relatively high proportion of
binaries ($f_v>0.8$) among systems that have $v<0.3$~km/s is obtained.
Significant differences between the binary proportions in the two
models can be found in the velocity intervals a) $-0.5<$~log$_{10}v<0$
and b) $0.1<$~log$_{10}v<0.6$. In interval~(a), $f_v\approx0.9,0.6$
and in interval~(b), $f_v\approx0.05,0.15$, for $R_{0.5}=2.5,0.8$~pc,
respectively.  Of all systems that finally emerge from such an
aggregate, 4~per cent have a velocity $v>5$~km/s.  These have a
slightly larger binary proportion than for the $R_{0.5}=2.5$~pc case
discussed above.  Overall, for $R_{0.5}\ge 0.8$~pc, $f_v$ {\it
decreases} monotonically with increasing $v$, and escaping stars have a
low ($f_v<0.2$) binary proportion.

Star formation in denser aggregates ($R_{0.5}<0.8$~pc) leads to
significantly different behaviour of $f_v$ with $v$
(Fig.~\ref{fig:binprop1}).  For $R_{0.5}=0.25$~pc, $f_v\approx0.45$
for $v<1$~km/s, with a discontinuous decrease to $f_v\le0.2$ for
$v>1$~km/s.

Of special interest is the $R_{0.5}=0.08$~pc model. It represents most
closely the Trapezium cluster, because it has a comparable central
number density, half-mass radius and velocity dispersion. In the
present model, the initial crossing and relaxation times are $t_{\rm
cr}=1\times10^5$~yr and $t_{\rm rel}=3\times10^5$~yr, respectively
(Table~1 in Kroupa 1995a), whereas in the Trapezium cluster, $t_{\rm
cr}\approx4-12\times10^5$~yr and $t_{\rm rel}\approx0.7-3.7$~Myr
(Bonnell \& Kroupa 1998). The Trapezium cluster, however, is different
in that it contains 500--1000~stars with a mean mass of about
$0.6\,M_\odot$, and in that it is the core of the much more massive
and extended Orion Nebula Cluster (Hillenbrand \& Hartmann
1998). Also, it is not clear if the entire cluster is gravitationally
bound.  

After dissolution of this $R_{0.5}=0.08$~pc aggregate most systems
have $v<1$~km/s (Fig.~\ref{fig:binprop1}).  The proportion of binary
stars shows a rather complex dependence on $v$.  The binary proportion
ranges from $f_v\approx0.1$ for systems with $v\approx0.1-0.2$~km/s to
$f_v\approx0.6$ for $v\approx1$~km/s; $f_v$ thus {\it increases} with
$v$ for $v<1$~km/s.  The binary proportion shows a significant maximum
($f_v\approx 0.6$) near $v\approx1$~km/s, and remains approximately
constant at $f_v\approx0.25$ for $3$~km/s$\,<v<20$~km/s. In this
rather extreme model of star formation, 4.6~per cent of all systems
have $v>5$~km/s after aggregate dissolution.  The low value of $f_v$
for small $v$, and its rise with $v$, is due to efficient disruption
of binaries at an early dynamical age, when the ejected stars are
decelerated most effectively by the young deep potential well. A
fraction of the predominantly single stars with low velocity, is also
a decelerated part of the high-velocity tail in the aggregates with
$R\ge0.8$~pc. This can be inferred from Figs.~\ref{fig:orbit1}
and~\ref{fig:orbit2} (Section~3.4). Hardened binaries (log$_{10}P<4$)
are found with small velocities. Usually they are the result of
energetic three-body or binary-binary interactions causing ejection.

Thus, in a realistic embedded cluster with the same stellar mass, the
initial escape velocity is larger because the potential is dominated
by the gas. Within a few~Myr most of the gas is expelled, leaving an
expanding cluster population, and a binary deficient remnant
population, in which each system has a small centre-of-mass velocity.
This decelerated and binary deficient population remains bound to the
molecular cloud and, after a few Myr, contributes to a distributed
population of young stars with significantly different dynamical
properties to the distributed population in Taurus-Auriga. Dispersal
of this binary deficient population takes long, and an observer finds
a loosely distributed group of stars of similar age, and with a
reduced binary proportion that depends on the initial cluster
concentration.
 
Concerning the aggregates with initially no primordial binaries
(Fig.~\ref{fig:binprop2}), more binaries form by capture in the
initially more concentrated aggregate ($R_{0.5}=0.08$~pc), owing to
the more frequent three-body encounters in the young concentrated
aggregate.  The resulting total binary proportion, however, remains
insignificant (Kroupa 1995a).  The data plotted in
Fig.~\ref{fig:binprop2} indicate that $f_v$ increases with $v$ (for
$v>1$~km/s), which is contrary behaviour to the aggregates that
contain a large population of primordial binaries.

\subsection{Mean system mass}
\label{subsec:meanmass}
\noindent
In a binary-binary or binary-single star encounter, binding energy can
be transformed into kinetic energy of the escaping stellar system. A
given acquired kinetic energy corresponds to a smaller ejection
velocity if the system mass is larger.  Low-mass stars can be ejected
with higher velocities than high-mass stars. The extensive simulations
performed by Leonard \& Duncan (1990) and Leonard (1991, 1995)
demonstrate that this is the case, and are consistent with the
observational mass-velocity diagram for OB stars produced by Gies \&
Bolton (1986). This is also discussed in length by Conlon et
al. (1990).

The present study concentrates on the mean-system-mass--velocity
relationship obtained from self-consistent $N$-body simulations of
clusters of low-mass stars ($m\le1.1\,M_\odot$), and is thus relevant
for the large-scale distribution of young-low mass stars seen in the
ROSAT survey (compare with Sterzik \& Durisen 1995).

As is evident from Figs.~\ref{fig:binprop1} and~\ref{fig:binprop2},
the behaviour of $<m>_v$ with $v$ is complex and depends on the
initial concentration of the aggregate. The complexities of the
underlying binary--binary and triple-star encounters are discussed at
length by Harrington (1975), Heggie (1975), Hills (1975), Leonard \&
Duncan (1990) and Leonard (1991, 1995), and a review can be found in
Valtonen \& Mikkola (1991). The most-apparent result here is that the
expected simple correlation (smaller $<m>_v$ for larger $v$) does not
hold, except for the $R_{0.5}=2.5$~pc aggregate. The few stars that
are ejected from this aggregate are low-mass stars expelled from
unstable triple or higher-order systems (Heggie 1975, Harrington 1975,
Hills 1977, compare also with the second-mass-family interactions of
Leonard 1991, and with Kiseleva, Eggleton \& Orlov 1994).

The expected correlation is also observed for $v>30$~km/s for all
binary-star aggregates. An acceleration beyond this velocity, which is
the orbital velocity of the hardest primordial binary in the present
simulations (Section~\ref{subsec:observables}), is destructive to all
primordial binaries. Only single stars appear with such large
velocities.

For $R_{0.5}=0.8, 0.25$ and 0.08~pc, the velocity range
$v\approx2$~km/s to~30~km/s yields no clear correlation between
$<m>_v$ and $v$: $<m>_v\approx0.5\,M_\odot$ is approximately
constant. This is an interesting finding which was also noted by
Harrington (1975). It shows that systems more massive than
$0.5\,M_\odot$ are ejected from aggregates similar to embedded
clusters, with velocities that can place them at distances between~20
and 300~pc within 10~Myr of ejection time from their formation
site. These systems result from stochastic and quite energetic
binary-binary and three-body encounters.

The majority of systems that have $v<1$~km/s, and which do not spread
much further than 1--10~pc from their formation site within 10~Myr,
have a constant $<m>_v$ that lies between the average stellar mass and
mean primordial system mass. Smaller values are seen for binary-poor
remnant populations (e.g. for $R_{0.5}=0.08$~pc). It follows that an
observer must be careful not to interpret the stellar mass function
and binary proportion of such populations in terms of a possible
dependence of these quantities on star-formation environment, without
due consideration of the dynamical history.

\subsection{Binding energy, period and mass ratio}
\label{subsec:ebin}
\noindent
Only binaries that are sufficiently bound will not be ionised when
they suffer a close interaction with another system, after which they
may leave the aggregate with relatively high velocity. Thus, the
correlation between binary star binding energy, $E_{\rm bin}$, and its
centre of mass kinetic energy, $E_{\rm kin}$, indicates the history of
a system.

Above it was seen that the final binary proportion, $f_v$, for systems
with 2~km/s$\,<v<30$~km/s, is larger for initially more concentrated
binary star aggregates. It can achieve values of 20-40~per cent for
$R_{0.5}\le0.25$~pc, although the overall final binary proportion is
small ($f_{\rm tot}\approx0.27$, fig.~3 in Kroupa 1995a). This result
is relevant for the distribution of young stars around star-forming
regions. Initially highly concentrated embedded clusters may add young
binaries to regions as far as 300~pc over a period of 10~Myr. Such
binaries have a well-defined correlation between $E_{\rm bin}$ and
$E_{\rm kin}$. This correlation transforms to correlations between $P$
and $v$, between the mass ratio ($q=m_2/m_1\le1$) and $v$, and 
between the system mass ($m_1+m_2$) and $v$, where $m_1$ and $m_2$ are
the primary- and secondary-star masses, respectively.

\subsubsection{Binding energy and period}
\label{sssec:eb_P}
\noindent
In Figs.~\ref{fig:orbit1} and~\ref{fig:orbit2} are plotted $E_{\rm
bin}$ against $E_{\rm kin}$, as well as the orbital period, $P$,
against ejection velocity, $v$, for each binary system in the $N_{\rm
run}$ simulations. The distribution of data points for the
$R_{0.5}=2.5$~pc aggregate (top two panels in Fig.~\ref{fig:orbit1})
reflects approximately the initial distribution in binding energies
(log$_{10}E_{\rm bin}>2$) and orbital periods (log$_{10}P\ge3$). Only
very few binaries have gained binding energy and/or have been
accelerated to higher velocities. As the initial $R_{0.5}$ is reduced,
the number of binary systems with hardened orbits and larger $v$
increases.  The depletion of orbits at large $P$ is clearly evident in
the $R_{0.5}=0.08$~pc cluster.

Figs.~\ref{fig:orbit1} and~\ref{fig:orbit2} nicely show that a binary
with an orbital period corresponding to $v_{\rm orb}$ only remains
bound when suffering a collision, if the ejection velocity $v<v_{\rm
orb}$. A single notable exception, that occurred in the five
simulations of the $R_{0.5}=0.08$~pc aggregate, is the binary system
with $P=10^{5.7}$~d and $v=16$~km/s. It is difficult to trace the
detailed dynamical history of any individual stellar system owing to
the discrete output times when stellar masses, positions and
velocities are written to computer disk. But this binary probably
formed in a complex high-order interaction, that resulted in two stars
being ejected on essentially identical trajectories.

The results discussed so far are valid for a primordial binary star
population that has periods $P>10^3$~d. In reality, binaries with
shorter periods do exist with a binary proportion $f\approx0.15$
(fig.~1 in Mathieu 1994). The inclusion of primordial binaries with
$P<10^3$~d does not change the results presented here, apart from
slightly increasing $f_v$ for the ejected systems. In
Figs.~\ref{fig:orbit1} and~\ref{fig:orbit2}, both the ($E_{\rm
bin},E_{\rm kin}$) and (log$_{10}P,v$) plots would contain orbits with
$E_{\rm bin}>10^2\,M_\odot$~km$^2$/s$^2$ and $P<10^3$~d, and an upper
envelope for $v>1$~km/s given by the dashed diagonal lines.

The correlation between binding and kinetic energy is established
particularly well for the binary systems that are formed by capture in
the single star aggregates (Fig.~\ref{fig:orbit3}). Their periods
range from $P>100$~d to $10^{11}$~d. The binary systems with
$P\approx10^{10}$~d form by chance three-body low-velocity encounters
and form a distinct group in the figure.

\subsubsection{Mass ratio and system mass}
\label{sssec:q_msys}
\noindent
More massive systems have a higher binding energy, and are thus more
resistant to higher accelerations. Binaries that are ejected from an
aggregate should thus have a larger system mass and a mass-ratio
nearer to unity for higher $v$. 

In Figs.~\ref{fig:mass1} and~\ref{fig:mass2}, $m_1+m_2$ and $q$ are
plotted in dependence of $v$ for the binary-star aggregates.  The
$R_{0.5}=2.5$~pc aggregate shows approximately the initial
distributions, and is useful as a reference. As the aggregate
concentration is reduced, the number of ejected systems increases.
These tend to have larger system masses, as expected. At the same
time, binary stars with a mass-ratio $q<0.2$ are preferentially
removed as $R_{0.5}$ is reduced.  This is expected because, for a
given semi-major axis, they have the lowest binding energy, $E_{\rm
bin}\propto m_1^2 \times q$.  The correlation between $q$ and $v$ is
in the expected sense, which is evident in the figures by the
appearance of orbits in the region $q>0.6, v>2$~km/s. However, the
correlation is weak, because $E_{\rm bin}$ also depends on $m_1$. As
is evident from the figures, the number of binaries with
$m_1+m_2>1.5\,M_\odot$ increases for smaller $R_{0.5}$. This is
because the more frequent three-body and higher-order interactions
lead to more frequent exchanges of companions, which most often leads
to the production of binary systems consisting of the most massive
stars involved in the interaction (Harrington 1975, Heggie 1975, Hills
1977, see also Valtonen \& Mikkola 1991, McDonald \& Clarke 1993).

Concerning the single-star aggregates, Fig.~\ref{fig:mass3} shows that
the distribution of mass ratios of the dynamically formed binaries in
the $R_{0.5}=0.25$~pc aggregate is roughly flat over the whole
accessible range. In the more concentrated aggregate, binaries form
with larger system mass and a bias towards larger $q$, which is a
result of dynamical biasing discussed in greater detail by McDonald \&
Clarke (1993).

\section{Conclusions}
\label{sec:conclusions}
\noindent
The correlations between ejection velocity and the proportion of
binaries, as well as their orbital parameters, have been quantified
for a range of initial dynamical configurations.  The correlations are
useful in the study of stellar systems that are apparently ejected
from Galactic clusters (see e.g. Frink et al. 1997), some of which are
known to be rich in binaries (e.g. Raboud \& Mermilliod 1998a,
1998b). Observed ejected binaries should show correlations as
presented in Figs.~\ref{fig:orbit1}, \ref{fig:orbit2},
~\ref{fig:mass1} and~\ref{fig:mass2}.  The results for the binary-rich
aggregates modelled here are also relevant for an understanding of the
large-scale distribution of young stars, because most stars appear to
form in aggregates with a high binary proportion.  Additionally, the
correlations contain information about the dynamical configuration at
birth.

For binary-rich aggregates containing a few hundred stars the
following correlations result: (i) more tightly clustered aggregates
lead to more stellar systems having larger ejection velocities and a
smaller overall binary proportion, (ii) the large population of
primordial binaries leads to a significantly enhanced number of
systems with high-ejection velocities compared to single-star
aggregates, (iii) systems with high ejection velocities have a
significantly reduced binary proportion, (iv) binary stars with high
ejection velocities have short-period orbits, and tend to be more
massive with a mass-ratio biased towards unity, (v) the average system
mass as a function of ejection velocity is defined above about 2~km/s
by stochastic close encounters, so that systems more massive than
$0.5\,M_\odot$ with high ejection velocities occur, and (vi)
aggregates with $R_{0.5}\le0.25$~pc lead to a complex dependence of
the resulting binary proportion on velocity, whereas a stellar
population emerging from less concentrated aggregates shows a
monotonic decrease of the binary proportion with increasing velocity.

For aggregates of a few hundred single stars one obtains: (i) more
tightly clustered aggregates lead to an increased number of stellar
systems with larger ejection velocities (but significantly less so
than in the binary-rich aggregates), and an enhanced overall binary
proportion that remains significantly below the observed binary
proportion in the Galactic field, (ii) the binary proportion increases
with ejection velocity, (iii) is as (iv) above, and (iv) is as (v)
above.

Remnant unbound young populations take long to disperse because they
have a small velocity dispersion. The binary proportion and mean
system mass (and thus the inferred IMF) of such a remnant population,
sensitively depends on the initial dynamical configuration of the
binary-rich birth aggregate. After emerging from the birth aggregate,
the distribution of velocities of a young stellar population changes
with time in the gravitational potential of the nearby molecular
clouds.  A substantial proportion of emerging stars is likely to
remain bound to the parent molecular cloud until it ceases to exist.

\begin{figure}[]
\plotfiddle{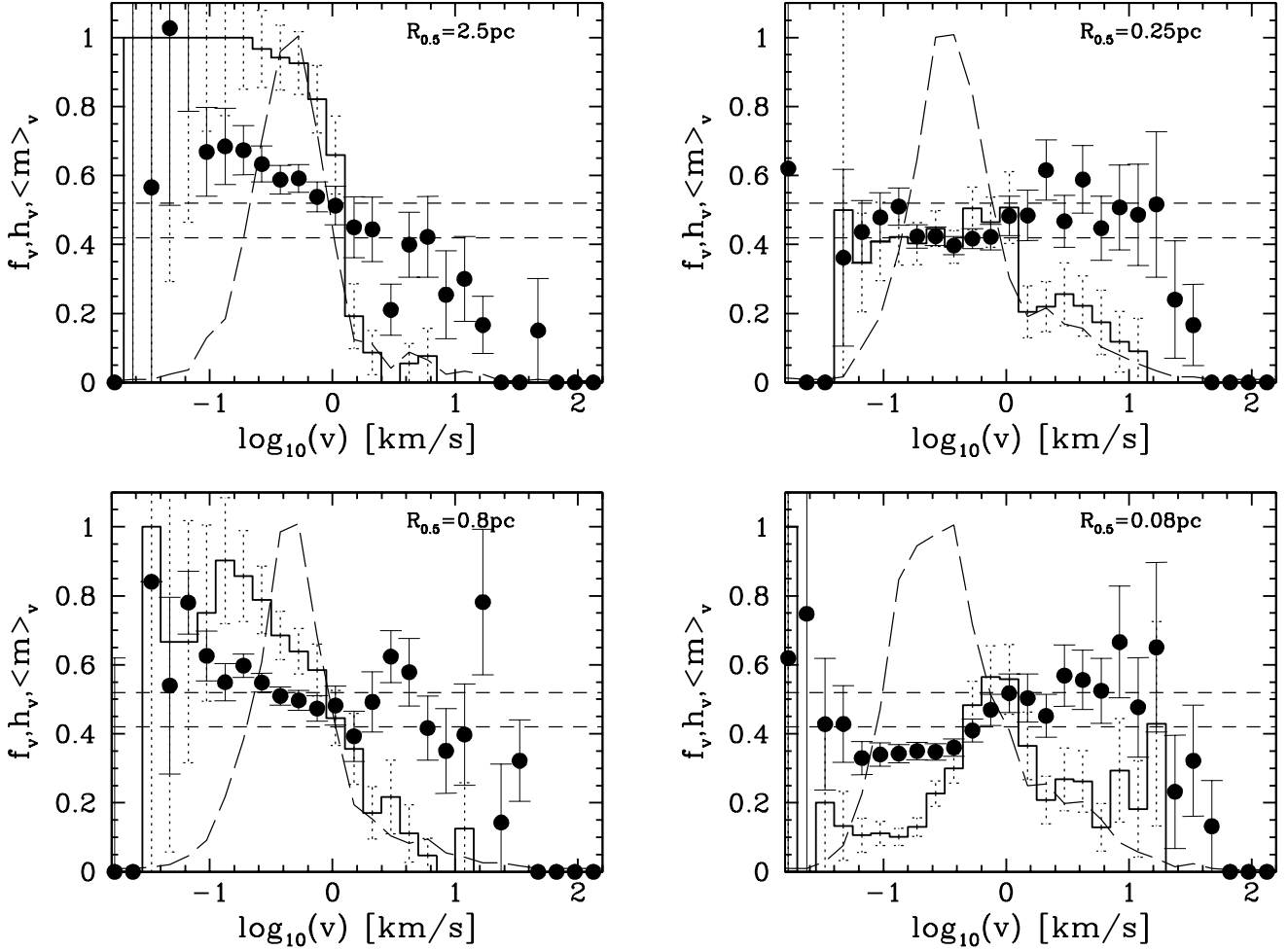}{15cm}{270}{70}{70}{-280}{430}
\caption{
\label{fig:binprop1}
Binary proportion, $f_v$, as a function of velocity, $v$, for the
aggregates with $R_{0.5}=2.5, 0.77, 0.25$ and 0.077~pc (solid
histogram). Solid circles are the mean system mass, $<m>_v$.
Error-bars are Poisson uncertainties.  Long-dashed curves represent
the distribution of velocities, $h_v$, plotted in arbitrary ordinate
units and tabulated in Table~1.  Horizontal dashed lines mark the
rough one-sigma range of the binary proportion for G-, K-, and M-main
sequence systems in the Galactic field ($f_{\rm tot}^{\rm
obs}=0.47\pm0.05$, Kroupa 1995a). From the value of $f_v$ around the
maximum of $h_v$, as well as the shape of the period and mass-ratio
distributions, it is deduced that 0.25~pc$\,<R_{0.5}<0.8$~pc are
solutions to inverse dynamical population synthesis (Kroupa 1995a).  }
\end{figure}

\begin{figure}[]
\plotfiddle{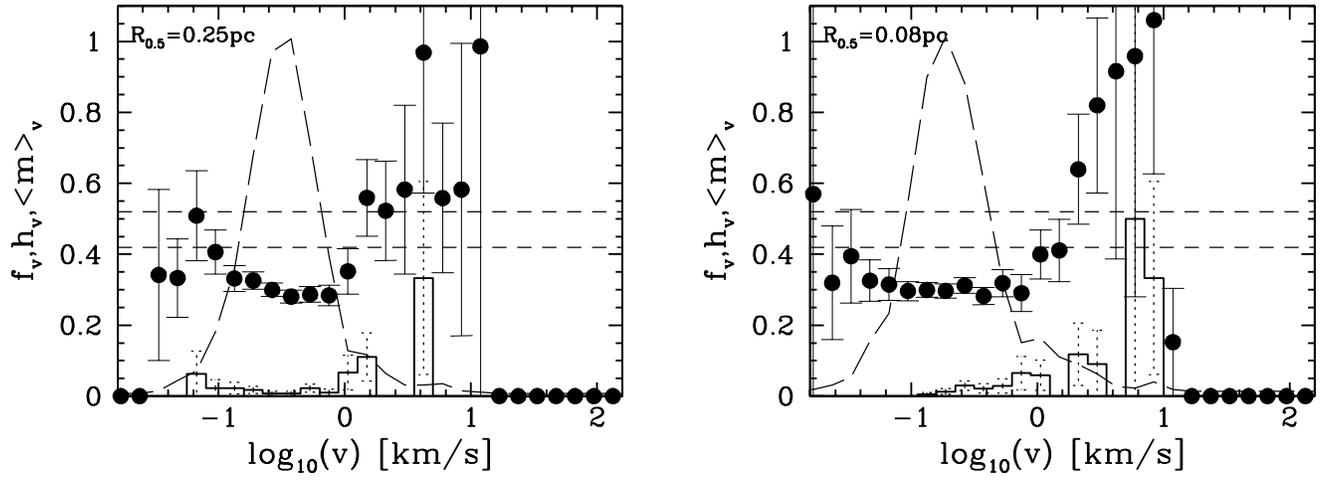}{15cm}{270}{70}{70}{-280}{430}
\caption{
\label{fig:binprop2}
As Fig.~\ref{fig:binprop1} but for stellar aggregates that initially
contain no binary system.  }
\end{figure}

\begin{figure}[]
\epsscale{0.8}
\plotone{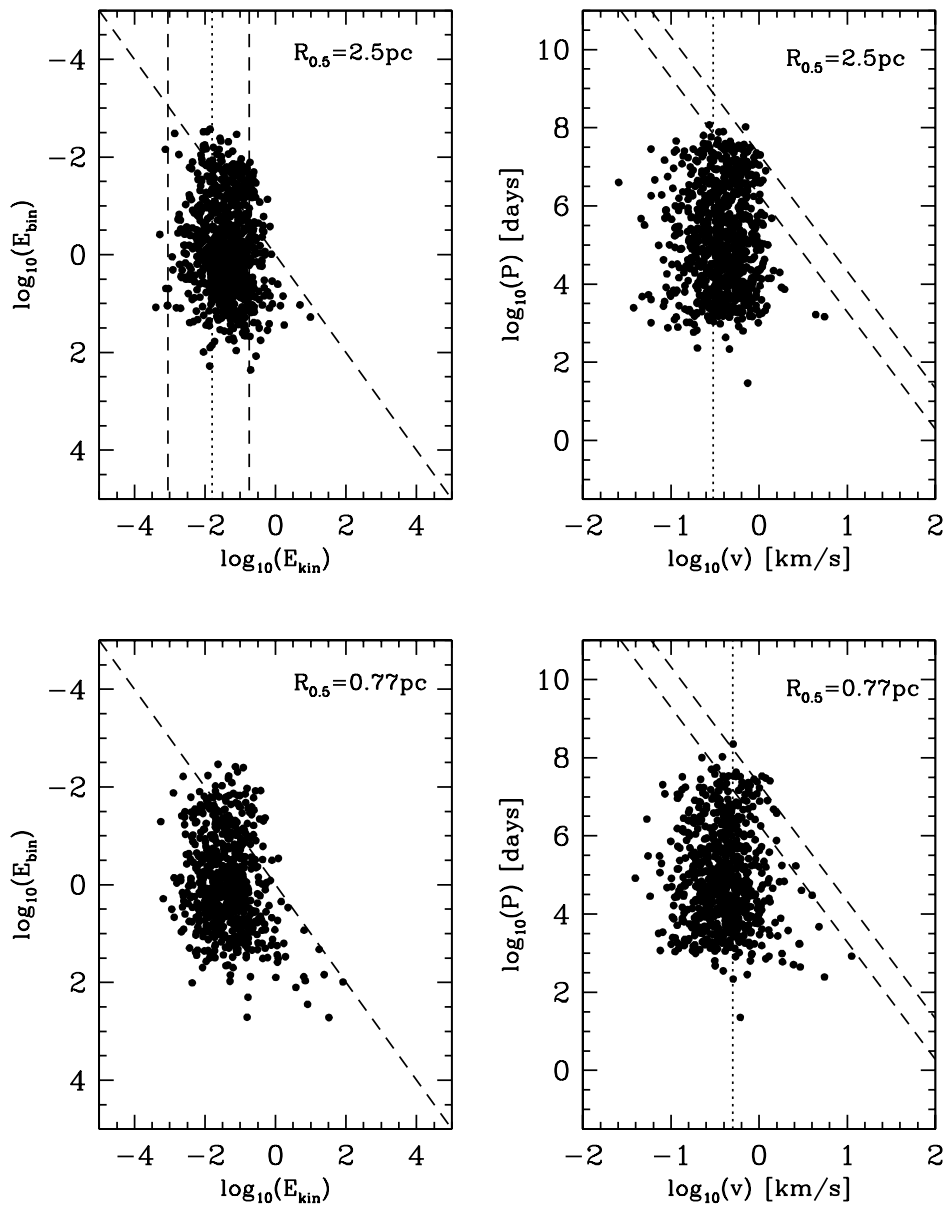}
\caption{
\label{fig:orbit1}
Distribution of orbits in the binding-energy--kinetic-energy plane
(left panels) (energy units are $M_\odot$ km$^2$/s$^2$) and the
period--velocity plane (right panels) for the two binary-star
aggregates with $R_{0.5}=2.5$~pc and~0.77~pc.  Orbits along the
diagonal dashed line have a binding energy that is equal to the
kinetic energy. The region between the two vertical dashed lines for
$R_{0.5}=2.5$~pc and $R_{0.5}=0.08$~pc (Fig.~\ref{fig:orbit2})
contains 95~per cent of all initial kinetic energies.  The vertical
dotted line plotted for these two aggregates approximates ${\overline
E_{\rm kin}}$.  Orbits with $E_{\rm bin}/{\overline E_{\rm kin}}>1$
are termed {\it hard} and are less likely to be ionised. Dashed lines
in the right panels are equation~4 for $m_{\rm sys}=2.2\,M_\odot$
(upper line) and $m_{\rm sys}=0.2\,M_\odot$ (lower line). The vertical
dotted lines are the initial velocity dispersions.  }
\end{figure}

\begin{figure}[]
\epsscale{0.8}
\plotone{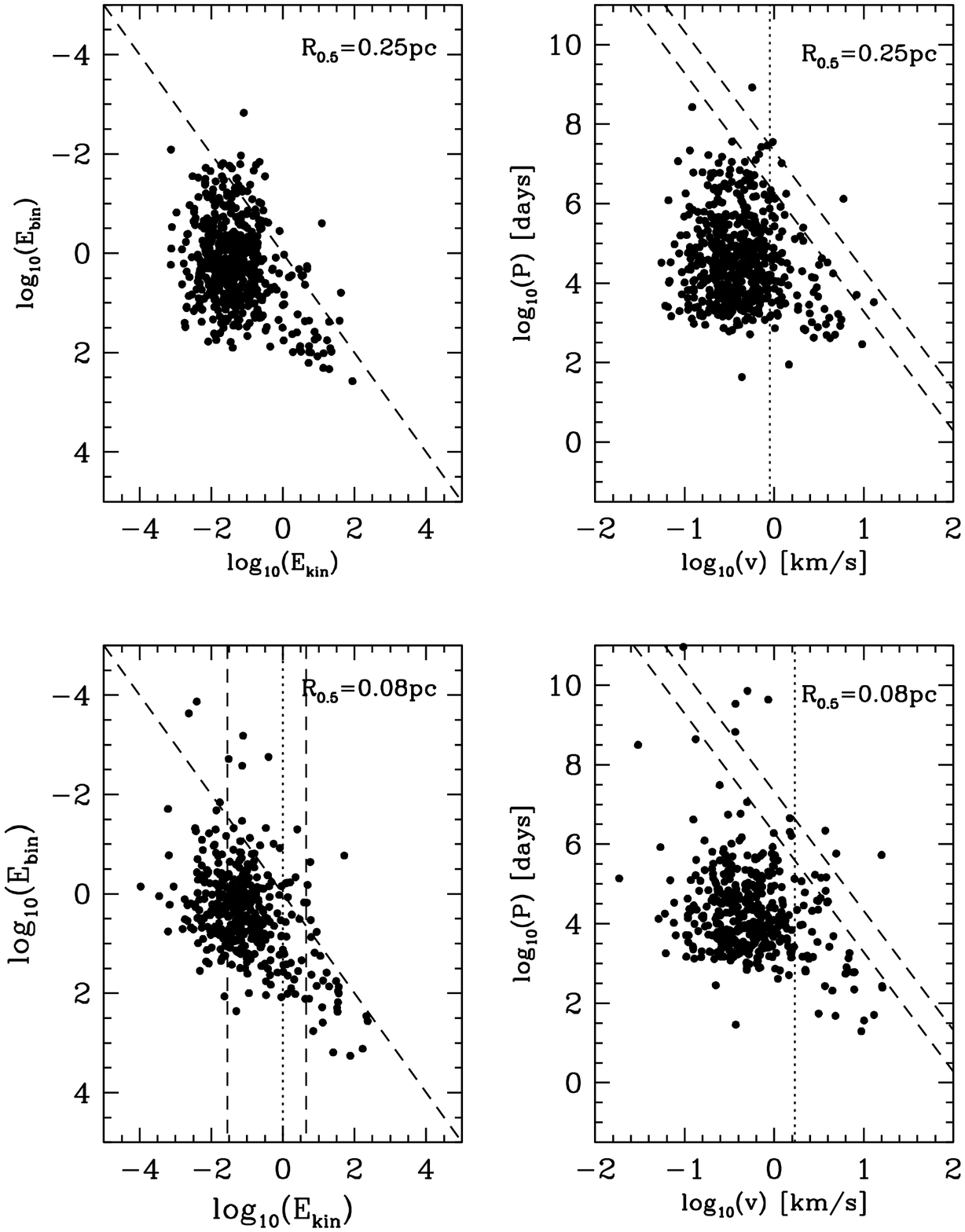}
\caption{\label{fig:orbit2} 
As Fig.~\ref{fig:orbit1} but for the two
binary-star aggregates with $R_{0.5}=0.25$~pc and~0.08~pc.  }
\end{figure}

These findings are important for interpreting the spatial
distribution, kinematics and binarity of young stars within and
surrounding star-forming regions.  Molecular clouds, in which stars
form preferentially in dense embedded binary-rich clusters, should
have an enhanced halo population of ejected and relatively binary poor
($f\approx0.25$) young stellar systems.  Also, young but
binary-depleted groups of stars can be misinterpreted to be evidence
for an environmental dependency of the binary-formation mechanism.
For example, in fig.~6 of Brandner et al. (1996), the region US-B has
more binaries than the region US-A, which also contains many more
B~stars than US-B. The presence of B~stars suggests that the stars in
US-A may have formed in dense embedded clusters. The stars seen in
US-A would then constitute the $v\simless1$~km/s remnant population
for $R_{0.5}\simless0.25$~pc (Fig.~\ref{fig:binprop1}).  Given the
results of the present study, it is suggested that such a difference
in binary proportion between two regions may be due to different
initial dynamical configurations, and need not imply a dependence of
the binary proportion on the star-forming environment.

\begin{figure}[]
\epsscale{0.8}
\plotone{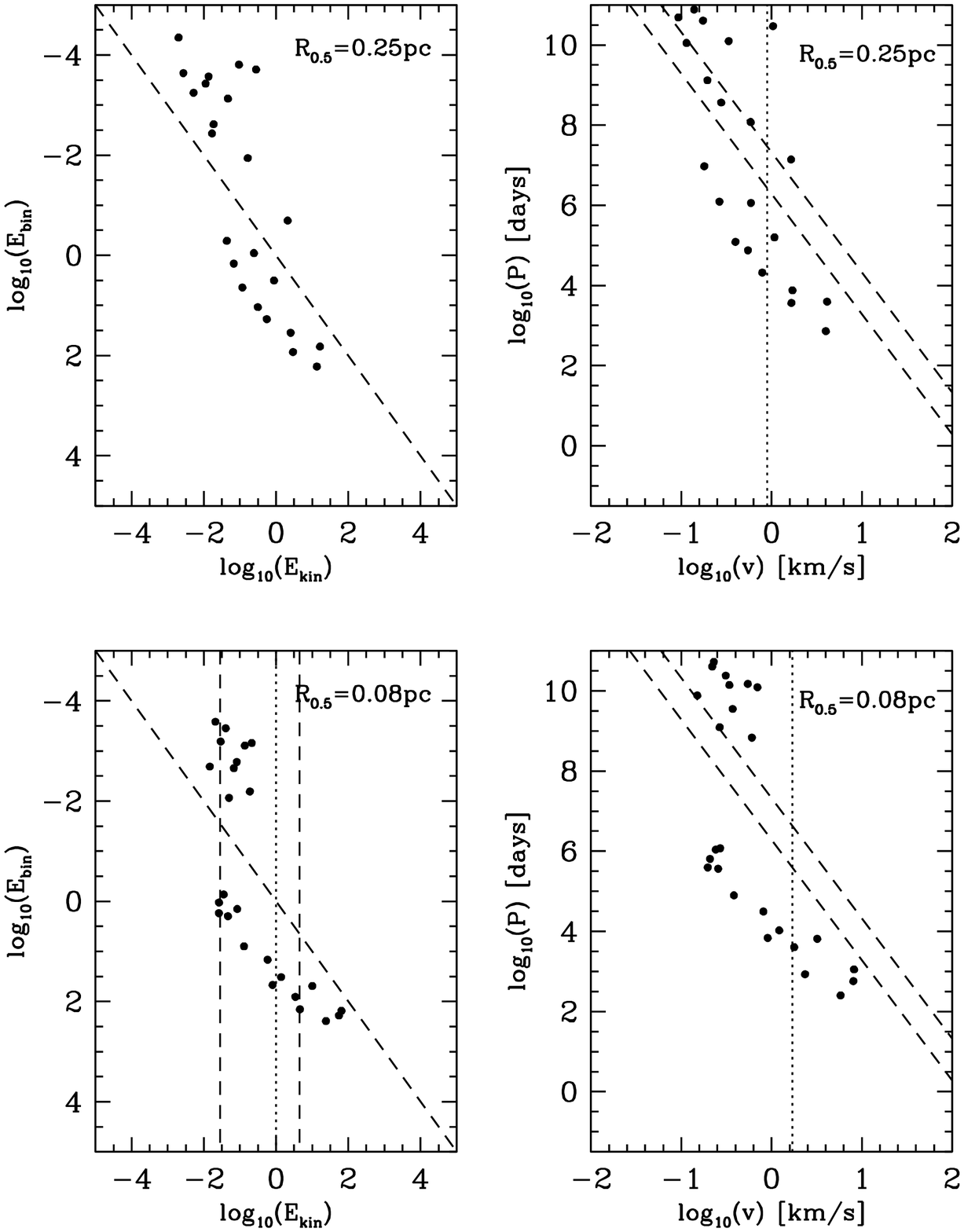}
\caption{\label{fig:orbit3}
As Fig.~\ref{fig:orbit1} but for the two single-star 
aggregates with $R_{0.5}=0.25$~pc and~0.08~pc.
}
\end{figure}

\begin{figure}[]
\epsscale{0.8}
\plotone{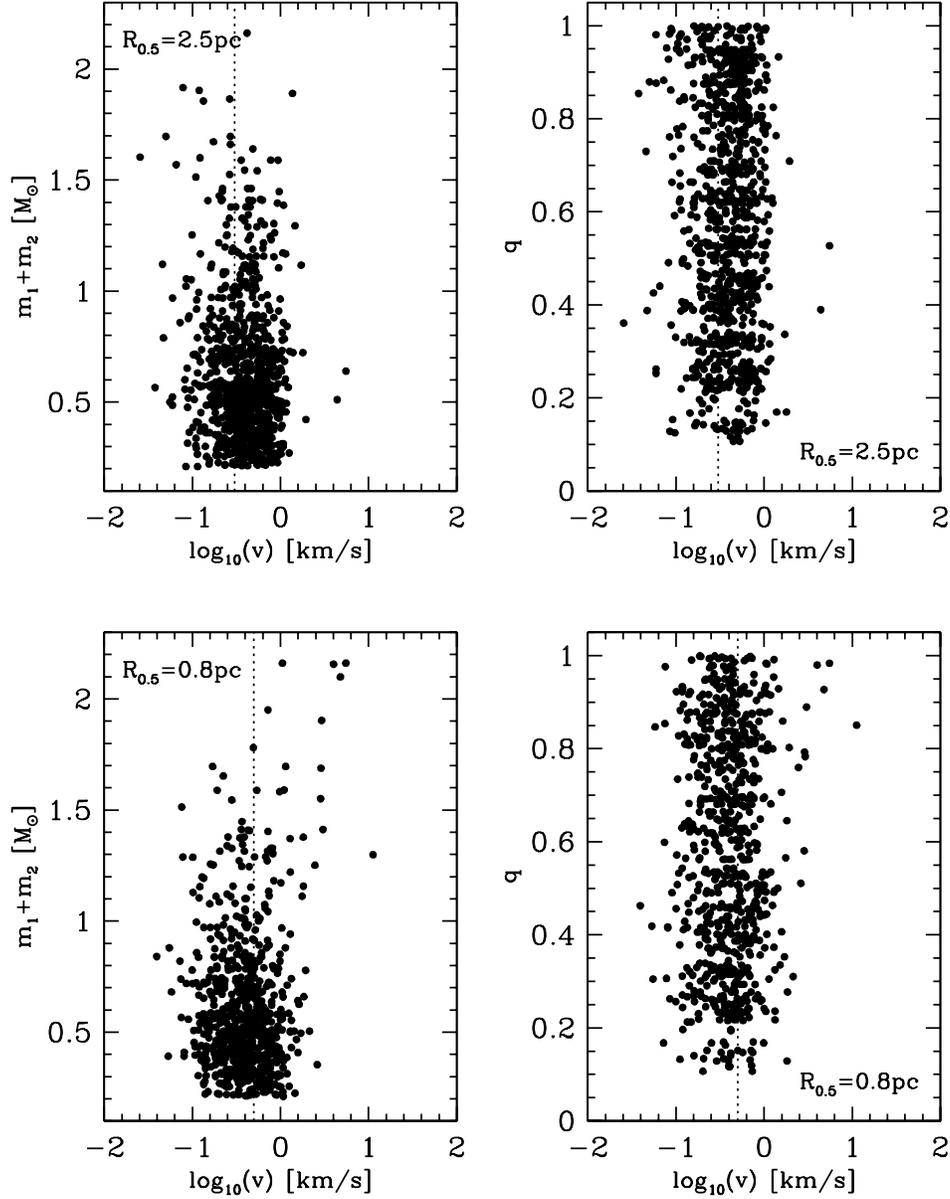}
\caption{\label{fig:mass1} Distribution of orbits in the
system-mass--velocity plane (left panels) and the mass-ratio--velocity
plane (right panels) for the two binary-star aggregates with
$R_{0.5}=2.5$~pc and~0.77~pc. The system mass is $m_1+m_2$ and the
mass ratio is $q=m_2/m_1\le1$, where $m_1$ and $m_2$ are the primary-
and secondary star masses, respectively. The vertical dotted lines
indicate the initial velocity dispersion.  }
\end{figure}

\begin{figure}[]
\epsscale{0.8}
\plotone{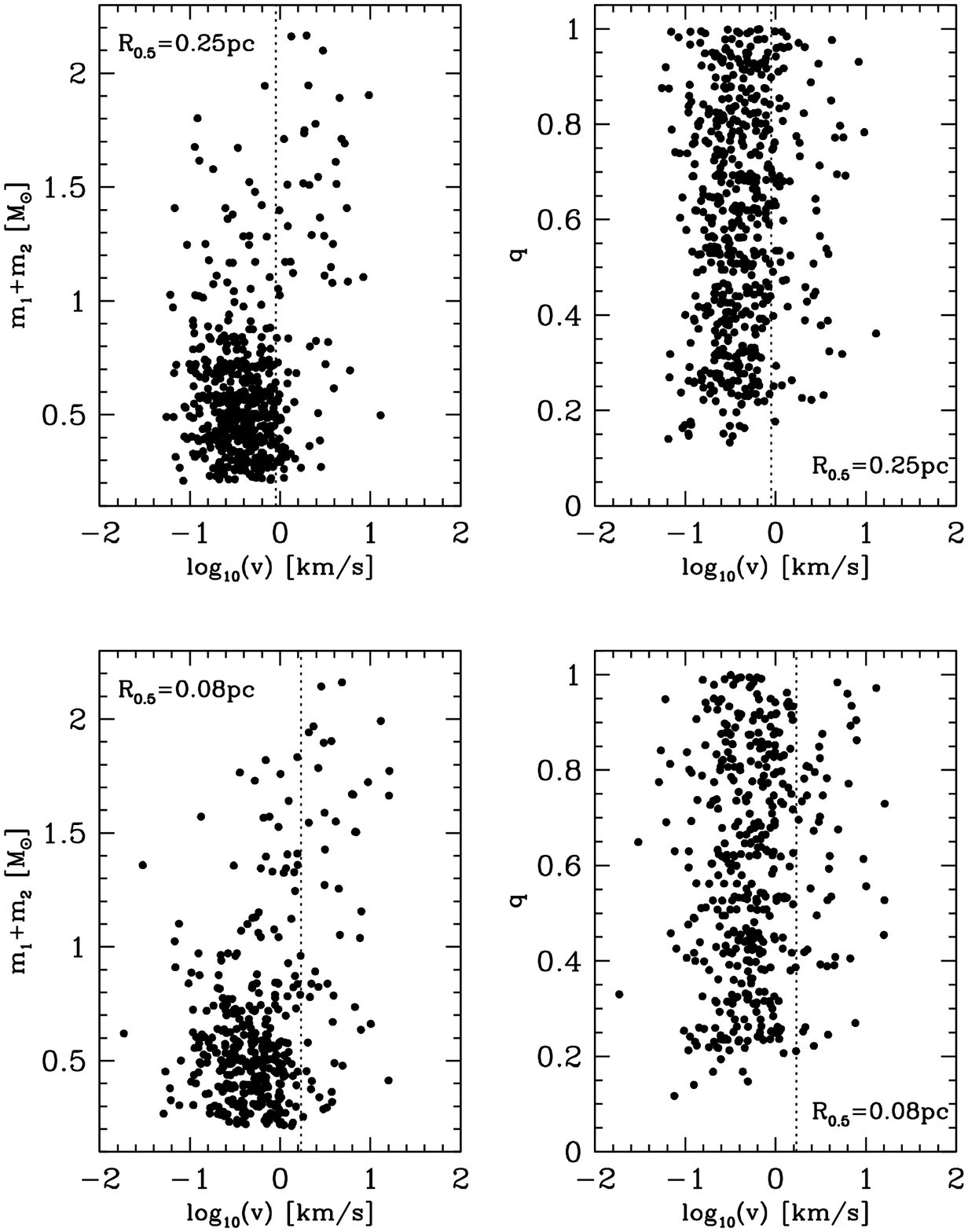}
\caption{\label{fig:mass2}
As Fig.~\ref{fig:mass1} but for the two binary-star 
aggregates with $R_{0.5}=0.25$~pc and~0.08~pc.
}
\end{figure}

\begin{figure}[]
\epsscale{0.8}
\plotone{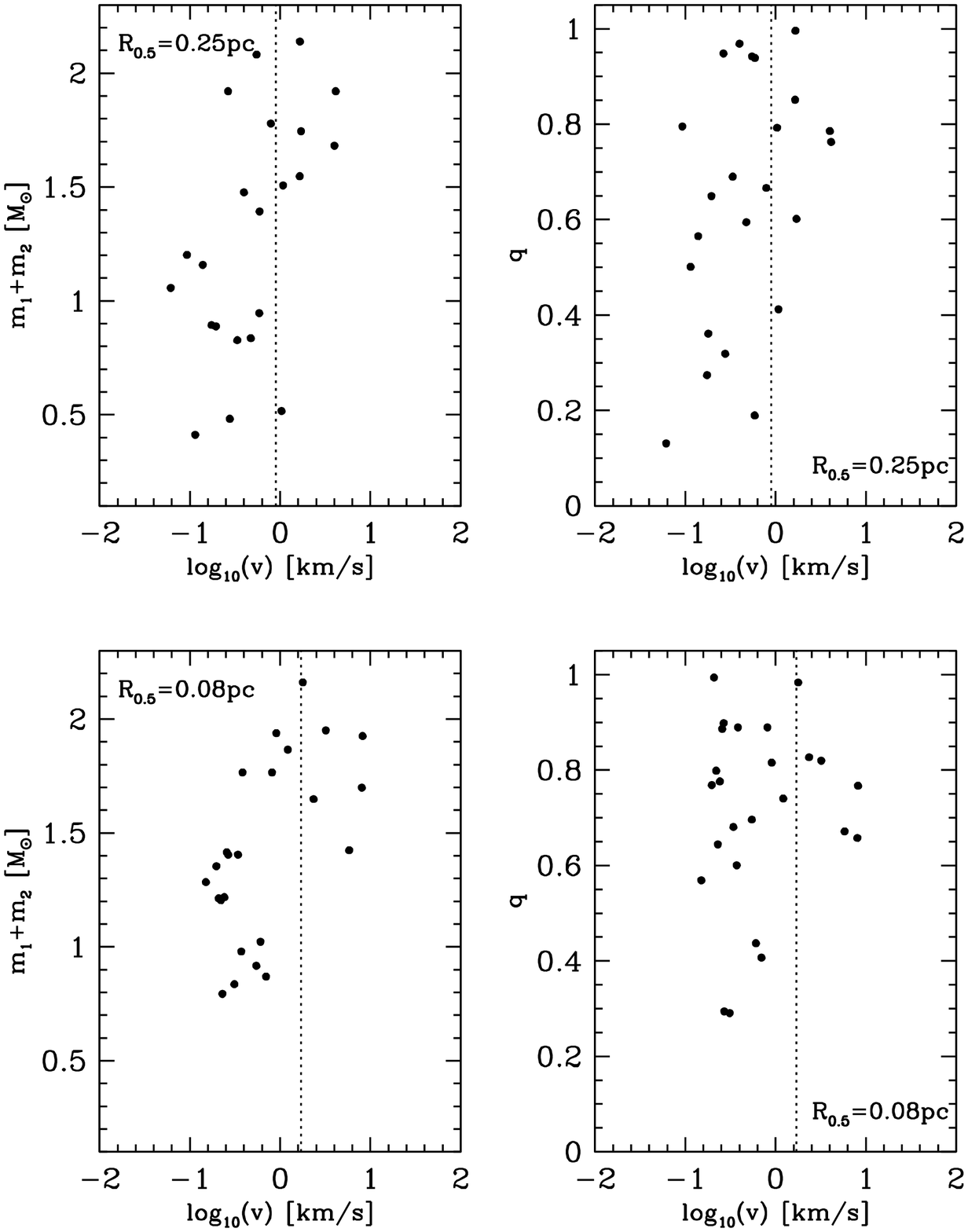}
\caption{\label{fig:mass3}
As Fig.~\ref{fig:mass1} but for the two single-star 
aggregates with $R_{0.5}=0.25$~pc and~0.08~pc.
}
\end{figure}

Important for the interpretation of the large-scale distribution of
young stars surrounding star forming sites is the realisation that
relatively massive systems are ejected with relatively large velocity
(2--30~km/s, Fig.~\ref{fig:binprop1}), which is a point also stressed
by Sterzik \& Durisen (1995).  The X-ray surveys are flux limited and
detect the massive stars (Wichmann et al. 1996), the presence of which
around star-forming regions may be a natural consequence of the
processes studied here.  However, if some young binary systems are
found to have orbital periods that place them above the dashed lines
in the right panels of Figs.~\ref{fig:orbit1}--\ref{fig:orbit3}, then
this would support the suggestion by Feigelson (1996), that
some star-formation occurs in small high-velocity clouds.


\acknowledgements 
\vskip 10mm
\noindent{\bf Acknowledgements}
\vskip 3mm
\noindent
I am very grateful to Sverre Aarseth for allowing me to use his NBODY5
programme, and I thank Rainer Spurzem and Mirek Giersz for helpful
discussions.


\clearpage


\begin{table}
{\small
\begin{minipage}[t]{20cm}
\hspace{-0.1cm}
  \begin{tabular}{*{8}{c}}
   \tableline\tableline
    log$_{10}v$~[km/s]  &&&~~~~~~~~~~~~~~$h_v$ [per~cent]\\
    \tableline
    $R_{0.5}[{\rm pc}]=$ &2.5   &0.8   &0.25   &0.08   
    &&$0.25^*$   &0$.08^*$ \\
    \tableline\tableline
$-1.925$ &  0    &  0   &  0    &  0.13 &&  0    &  0.09 \\
$-1.775$ &  0    &  0   &  0.07 &  0.06 &&  0    &  0.09 \\
$-1.625$ &  0.09 &  0   &  0    &  0.06 &&  0    &  0.34 \\
$-1.475$ &  0.09 &  0.08&  0    &  0.32 &&  0.17 &  0.77 \\
$-1.325$ &  0.37 &  0.24&  0.14 &  0.95 &&  0.77 &  2.73 \\
$-1.175$ &  0.64 &  0.73&  1.61 &  2.98 &&  1.36 &  4.26 \\
$-1.025$ &  2.48 &  1.61&  3.09 &  6.29 &&  3.66 & 10.22 \\
$-0.875$ &  3.59 &  4.12&  6.31 & 11.37 &&  7.40 & 17.12 \\
$-0.725$ &  8.19 &  7.34& 10.80 & 12.70 && 14.20 & 19.34 \\
$-0.575$ & 13.89 & 11.78& 16.83 & 13.14 && 20.32 & 16.78 \\
$-0.425$ & 19.04 & 19.21& 16.97 & 13.52 && 21.09 & 11.50 \\
$-0.275$ & 19.96 & 19.69& 14.03 &  9.59 && 14.97 &  5.96 \\
$-0.125$ & 14.35 & 13.24&  9.05 &  6.86 &&  8.16 &  2.64 \\
$+0.025$ &  7.82 &  8.15&  4.98 &  5.46 &&  2.55 &  2.90 \\
$+0.175$ &  2.39 &  3.63&  3.09 &  3.30 &&  2.30 &  1.87 \\
$+0.325$ &  2.12 &  2.82&  3.51 &  3.37 &&  1.19 &  1.45 \\
$+0.475$ &  0.74 &  1.86&  2.73 &  2.60 &&  0.51 &  0.94 \\
$+0.625$ &  1.66 &  1.45&  2.52 &  2.67 &&  0.51 &  0.26 \\
$+0.775$ &  1.20 &  1.69&  1.61 &  1.97 &&  0.60 &  0.17 \\
$+0.925$ &  0.37 &  0.89&  1.19 &  1.08 &&  0.17 &  0.51 \\
$+1.075$ &  0.55 &  0.65&  0.77 &  0.70 &&  0.09 &  0.09 \\
$+1.225$ &  0.37 &  0.32&  0.42 &  0.44 &&  0    &  0    \\
$+1.375$ &  0    &  0.32&  0.14 &  0.13 &&  0    &  0    \\
$+1.525$ &  0    &  0.16&  0.14 &  0.25 &&  0    &  0    \\
$+1.675$ &  0.09 &  0   &  0    &  0.06 &&  0    &  0    \\
      \tableline\tableline
\end{tabular}
\end{minipage}
}
\caption{\label{table:vdist}
Velocity distributions. $R_{0.5}^*$ is for the two single-star
aggregates. }
\end{table}

\clearpage
\newpage


\end{document}